\documentclass[12pt,preprint]{aastex}
\usepackage{graphicx}
\usepackage{amsmath}

\newcommand{\myemail}{Pierre-Yves.Longaretti@obs.ujf-grenoble.fr}

\slugcomment{Accepted for publication in ApJ}

\shorttitle{Shear Turbulence Phenomenology}
\shortauthors{Longaretti}

\begin{document}

\title{On the Phenomenology of Hydrodynamic Shear Turbulence}

\author{Pierre-Yves Longaretti}
\affil{Laboratoire d'Astrophysique de Grenoble, BP 53X, Grenoble
Cedex 9, 38410, France}

\email{\myemail}

\begin{abstract}
The question of a purely hydrodynamic origin of turbulence in
accretion disks is reexamined, on the basis of a large body of
experimental and numerical evidence on various subcritical (i.e.,
linearly stable) hydrodynamic flows.

One of the main points of this paper is that the length scale and
velocity fluctuation amplitude which are characteristic of
turbulent transport in these flows scale like $Re_m^{-1/2}$, where
$Re_m$ is the minimal Reynolds number for the onset of fully
developed turbulence. From this scaling, a simple explanation of
the dependence of $Re_m$ with relative gap width in subcritical
Couette-Taylor flows is developed. It is also argued that flows in
the shearing sheet limit should be turbulent, and that the lack of
turbulence in all such simulations performed to date is most
likely due to a lack of resolution, as a consequence of the effect
of the Coriolis force on the large scale fluctuations of turbulent
flows.

These results imply that accretion flows should be turbulent
through hydrodynamic processes. If this is the case, the
Shakura-Sunyaev $\alpha$ parameter is constrained to lie in the
range $10^{-3}-10^{-1}$ in accretion disks, depending on unknown
features of the mechanism which sustains turbulence. Whether the
hydrodynamic source of turbulence is more efficient than the MHD
one where present is an open question.
\end{abstract}

\keywords{hydrodynamics -- turbulence -- accretion disks}

\section{Introduction}

The need for turbulent transport to account for the rather short
accretion/ejection time-scales of Young Stellar Objects (YSOs) and
binary systems (CV, X-ray binaries), or for the very large energy
output of Active Galactic Nuclei (AGNs), is a well-known feature
of accretion disk theory. From the very beginning, differential
rotation has been regarded as one of the most promising sources
for turbulence, since shear flows are known to be able to feed
both hydrodynamic and MHD instabilities.

In most accretion disk models, the angular velocity profile
satisfies Rayleigh's criterion, implying that the corresponding
hydrodynamic flow is linearly stable. This is the case in
particular of the nearly keplerian velocity profile of cold disk
models. However, finite amplitude instabilities are theoretically
known to occur in some linearly stable flows, and are believed to
cause the turbulence observed in actual experiments, e.g. in
planar Couette flows, or Couette-Taylor flows with the inner
cylinder at rest. Furthermore, shear-driven hydrodynamic
turbulence would certainly produce the required outward transport
of angular momentum for keplerian flows, due to their outwardly
decreasing angular velocity profile. For these reasons, turbulence
in accretion disks (magnetized or not) has long been widely
believed to originate in purely hydrodynamic phenomena.

This picture has seriously been challenged in the past decade.
First, \citet{BH91} have shown that a local version of the
magneto-rotational instability \citep{Chandra60} operates in
differentially rotating disk. This instability was later
recognized to give rise to MHD turbulence and transport as well as
to magnetic field amplification \citep{HGB95, brand95}. The
physics of the magneto-rotational instability is by now a
well-established aspect of accretion disk theory. Secondly, recent
simulations of hydrodynamic fluid flows in the shearing sheet
approximation strongly suggest that accretion disk flows cannot
become turbulent through hydrodynamic processes alone
\citep{BHS96, HBW99}; indeed, in these simulations, planar Couette
flows are observed to be turbulent, but turbulence disappears as
soon as a Coriolis force is added, suggesting that this force
prevents the onset of the finite amplitude instabilities through
which linearly stable flows are believed to become turbulent.
Although this last finding seems to conflict with the available
experimental evidence on Couette-Taylor flows \citep{RZ99}, it has
strengthened the idea that linear magnetic instabilities play a
key role in the onset of turbulence in accretion disks.

The main objective of this paper is to critically reinvestigate
the possibility of hydrodynamic turbulent motions in accretion
disks, especially for linearly stable flows. A possible
hydrodynamic origin of turbulent motions is important for several
reasons. First, differential rotation is universally present in
disks, whereas some disks or disk regions might not be ionized
enough to support MHD phenomena, and a non-MHD source of
turbulence must be at work there. For instance, protoplanetary
disks are probably too resistive to support MHD turbulence
\citep{Flem00,Sano00}, but their observationally constrained
accretion rates imply the existence of turbulent transport.
Secondly, the existence of self-consistent magnetized
accretion/ejection structures seems to require a quasi
equipartition of thermal and magnetic energy which, combined with
the vertical stratification of these structures, may prevent the
development of the magneto-rotational instability in a number of
instances \citep{Ferr,Cassferr}.

The objectives of this paper are achieved through two different
means. First, relevant pieces of information on the behavior of
various types of shear flows which are available in the
specialized fluid dynamical literature are presented; from this
material, it is argued that shearing sheet flows should be
turbulent. Second, a phenomenological description and
understanding of relevant turbulent properties of shear flows as
they appear in the available experiments and numerical simulations
is developed. In its most basic form, the phenomenology of
hydrodynamic turbulent flows often relies on the concepts of
Kolmogorov cascade and turbulent viscosity, and this approach is
adopted here. Although the turbulent viscosity concept is of
limited validity in complex situations \citep{Tenlum} and its use
in the assessment of stability properties of turbulent flows has
been rightly criticized (see, e.g., \citealt{TER01};
\citealt{HBS01}), it is well-known to provide accurate scalings of
mean flow properties in simple shear flows such as channel or
planar Couette flows (\citealt{Tenlum,les}); consequently it has
been widely used to parameterize turbulent transport in accretion
disks. In this paper, some new and interesting consequences of
this ansatz and their implications for turbulence in shear flows
are pointed out.

This paper is organized as follows. In the next section, some
relevant flow configurations are introduced, along with the
related forms of the Navier-Stokes equations; I also summarize
relevant features of turbulence in these flows, as found in the
literature, as this material has some direct bearing to the
question of hydrodynamic turbulence in accretion disks, and is
largely unknown to the astrophysical community; the reader who is
not interested in factual details but only on their significance
can jump directly to section \ref{ssturb}, where this material is
used to infer that ``perfect" shearing sheet simulations should be
turbulent. The most interesting findings of the present work are
collected in section~\ref{turbvis}; after briefly recalling the
origin and rationale of the turbulent viscosity prescription, some
of its previously unnoticed but important consequences are derived
and used to interpret the behavior of the flows previously
described, with special attention paid to Couette-Taylor flows,
and to flow description in the shearing sheet approximation; in
particular, a phenomenological explanation of the scaling of the
Reynolds number with gap width in subcritical Couette-Taylor flows
is devised. On the basis of this phenomenological understanding,
the various reasons which might likely prevent the onset of
turbulence in the simulations of \citet{BHS96} and \citet{HBW99}
are discussed and the most critical one identified. The final
section summarizes the most relevant conclusions and discusses
their consequences for accretion disk theory and simulation, in
particular on the magnitude of the Shakura-Sunyaev $\alpha$
parameter.

The reader interested only in the new results of this paper and
not in the background fluid mechanical information can focus on
sections \ref{ssturb}, \ref{turbord} to \ref{cor}, and \ref{dis}.

\section{Turbulence in hydrodynamic shear flows} \label{phen}

Hydrodynamic accretion disk mean flows are widely believed to be
subcritical, i.e., the viscously relaxed laminar flow is linearly
stable at all Reynolds numbers, at least locally. Furthermore, all
experiments and numerical simulations of interest here pertain to
linearly stable flows, and we are mostly interested in the local
generation of turbulence. Therefore, I will focus on subcritical
flows in this paper.

The transition to turbulence is usually rather different in
subcritical and supercritical flows. Supercritical flows undergo a
cascade of precisely defined bifurcations in parameter space,
eventually leading to fully developed turbulence; these
transitions are well documented and reproduced numerically, e.g.
for Couette-Taylor flows (\citealt{And86} and references therein;
\citealt{Marc84a, Marc84b}). Turbulence in subcritical flows, on
the contrary, may abruptly be triggered, most probably by finite
amplitude instabilities \citep{Dauch94}; also, the flow apparently
evolves from highly intermittent to fully turbulent over a range
of Reynolds numbers.

Furthermore, shear flows can be either (wall-)bounded or free. The
distinction refers to the limitation of the flow in the direction
where the shear is applied (the transverse or shearwise
direction). This difference in boundary conditions influences some
of their turbulent properties; indeed, free flows are
characterized by a single length-scale, the extent of the shear
layer, whereas the distance to the wall introduces a second length
scale in wall-bounded flows. The influence of the other
(streamwise and spanwise) boundaries is minimized inasmuch as
their spacing exceeds the coherence length of the largest
turbulent eddies, and as globally induced perturbations (such as
Ekman circulation) are minimized by appropriate designs of the
experimental setups.

Shear flows have been actively studied in the past decades, and
their turbulent properties are now characterized for a large
variety of settings. In this section, I will briefly present the
subcritical flows which have direct bearing to the question of
hydrodynamic turbulence in accretion disks, namely, plane Couette
and free shear flows, either rotating or not, Couette-Taylor
flows, and Rayleigh-stable tidally driven shear flows in the
shearing sheet approximation. The first two have been studied
through both experiments and numerical simulations. On the
contrary, information on Rayleigh stable Taylor-Couette flows
comes only from experiments. Finally, the shearing sheet
approximation has been widely used as a local analytic model of
local accretion disks, and has been implemented in the numerical
work of Balbus, Hawley and coworkers quoted in the introduction
\citep{BHS96, HBW99}. For each of these flows, I characterize the
geometry, the critical parameters which are important for the
question of the onset of turbulence, and I also give the governing
dynamical equation (Navier-Stokes) in the form which is most
suitable to establish comparisons between the various types of
flows.

The object of this section is to try to give an answer to the
following question: if numerical simulations were perfect (i.e.,
not limited by questions of resolution, numerical instabilities
etc), would shear flows be turbulent in presence of the Coriolis
force ? This is done in section \ref{ssturb}, with the help of the
material collected here.

\subsection{Plane Couette and free shear flows}\label{plcou}

\begin{figure}
  \plotone{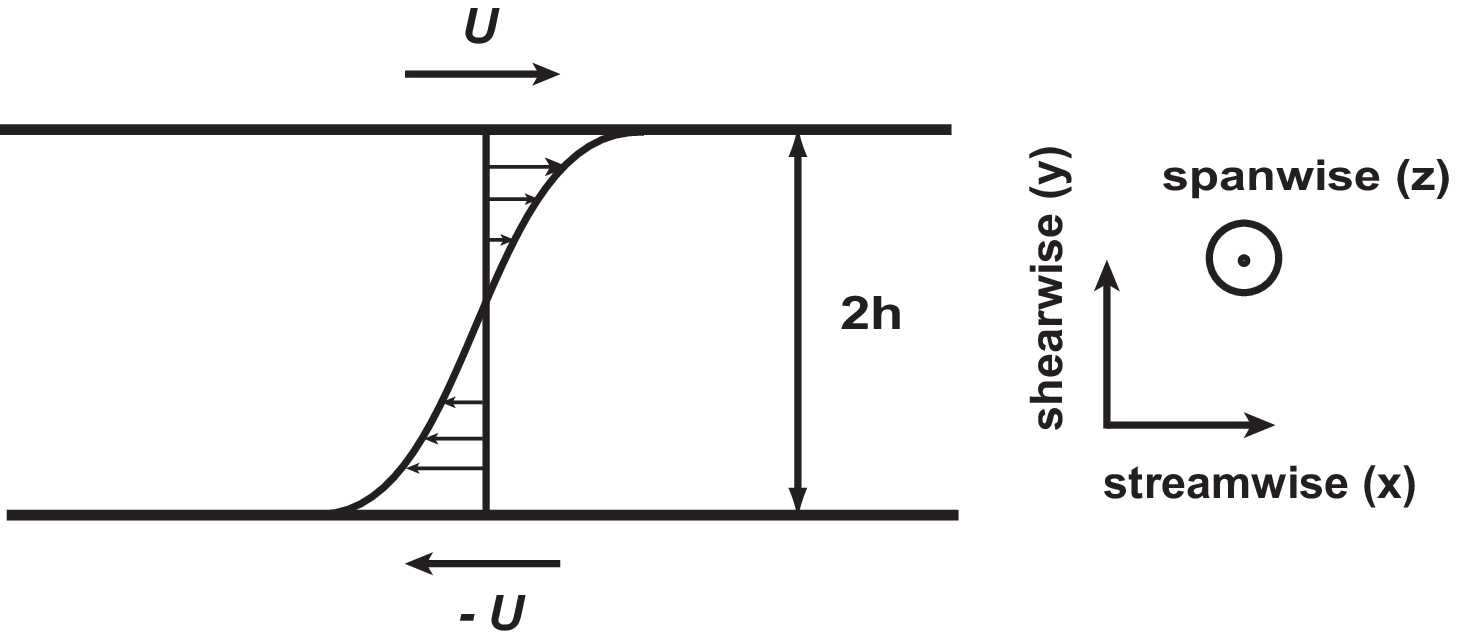}
  \caption{\small Sketch of the configuration of Couette
  flows. The flow is bounded by two counter-moving walls, and boundary layers develop in the
  turbulent regime, as shown by the mean velocity profile. By putting the experimental setup
  on a rotating platform, one obtains the so-called rotating Couette flow. }\label{fig1}
\end{figure}

\begin{figure}
  \plotone{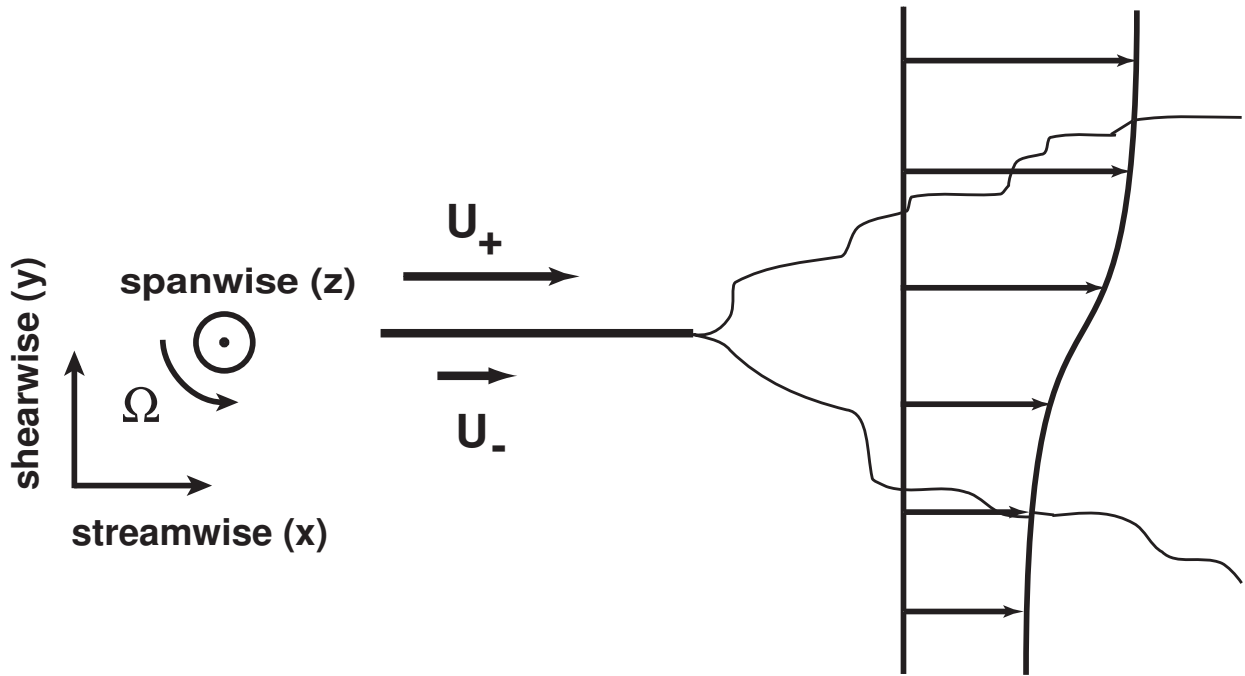}
  \caption{\small Sketch of the configuration of (rotating) free shear layers. Two layers of fluid
  of different velocities, initially horizontally separated, come in contact at the end of a
  dividing plate, and a turbulent shear layer develops and widens downstream.}\label{fig2}
\end{figure}

In spite of their conceptual simplicity, plane Couette flows are
difficult to produce in actual experiments, which explains why
some of their basic turbulent properties have only recently been
characterized. The experimental setup is schematically represented
on Fig.~\ref{fig1}, along with a sketch of the turbulent mean flow
profile (see \citealt{Till92} for details). In practice the two
walls are often made up of counter-moving (looped) infinite belts.
Similarly, free shear layers are produced by injecting fluid with
different velocities on each side of a separating plate. The
fluids come in contact at the end of the plate, and a turbulent
layer develops and widens downstream (see Fig.~\ref{fig2}).

These flows are described by the Navier-Stokes equation in its
simplest form, which reads

\begin{equation}\label{NSC}
  \frac{\partial {\bf v}}{\partial t}+{\bf v}.\nabla{\bf v}
  =-\frac{\nabla{P}}{\rho}+\nu\Delta{\bf v},
\end{equation}

\noindent with obvious notations. The viscous terms are displayed
in the incompressible form, as we are mostly concerned with
subsonic turbulence.

It is customary to define the Reynolds number of plane Couette
flows based on the half-velocity difference (i.e. $U$) and
half-width distance (i.e. $h$) between the two walls. However, for
the purpose of comparison with other setups, I shall define the
Reynolds number as

\begin{equation}\label{couetteRe}
  Re=4Uh/\nu,
\end{equation}

\noindent i.e. based on the total velocity difference and distance
between the two boundaries; the reader should bear in mind the
resulting factor of $4$ when comparing the figures quoted in this
paper for Couette flows with the literature. From the experiments
of \citet{Till92}, the minimal Reynolds for which turbulence is
sustained is $Re\simeq 1500$. The onset of turbulence in planar
Couette flow has been successfully reproduced in numerical
simulations (e.g., \citealt{Bech95} and references therein); a
nonlinear mechanism for tapping the mean shear to sustain
turbulence has even been identified (e.g., \citealt{Jim91, Ham95,
Wal97}).

Rotating Couette and rotating free shear flows are produced by
placing the experimental setup on a rotating platform. Such flows
are very relevant to astrophysics, as they share a number of
features with accretion disk flows in the shearing sheet
approximation. For these flows, the Navier-Stokes equation reads

\begin{equation}\label{NSCR}
  \frac{\partial{\bf v}}{\partial t}+{\bf v}.\nabla{\bf v}
  =-\frac{\nabla{P}}{\rho}-2{\bf \Omega}\times{\bf v}+{\bf F_{\rm in.}}+\nu\Delta{\bf
  v},
\end{equation}

\noindent where ${\bf F_{\rm in.}}$ stands for the inertial force
due to rotation.

 These rotating flows are usually simulated by including only the
Coriolis force term\footnote{The centrifugal term is not included
on the basis that it results only in a redistribution of the
equilibrium pressure.} in the Navier-Stokes equation
\citep{Bech96a, Komm96, Bech97}, so that Eq.~(\ref{NSCR}) reduces
to

\begin{equation}\label{NSR}
  \frac{\partial{\bf v}}{\partial t}+{\bf v}.\nabla{\bf v}
  =-\frac{\nabla{P}}{\rho}-2{\bf \Omega}\times{\bf v}+\nu\Delta{\bf v}.
\end{equation}

\noindent Rotating Couette and rotating free shear flows are
characterized by the ratio $S$ of the angular velocity of rotation
to the shear

\begin{equation}\label{Ro}
  S=-\frac{2\Omega}{d\langle v_x\rangle/dy},
\end{equation}

\noindent where $\langle v_x\rangle$ is the mean velocity profile;
this number is akin to an inverse Rossby number, and measures the
relative strength of the Coriolis and advection terms in the
Navier-Stokes equation. A linear shear is destabilized by rotation
when

\begin{equation}\label{Ro-uns}
  -1 < S < 0,
\end{equation}

\noindent and stabilized otherwise (see \citealt{Trit92} and
references therein; see also section \ref{ssturb}). The relevant
regime for astrophysics is $S\lesssim -1$ (i.e., negative $S$ and
linearly stable linear shear\footnote{In relating rotating flows
to shearing sheet ones, notice that the $y$ axis identifies to the
radial one, whereas the $x$ and azimuthal directions are
antiparallel.}). No systematic exploration of the ($Re$, $S$)
parameter space has been performed; furthermore, I am not aware of
any experimental investigation of rotating Couette flows for such
(relatively) high values of $S$. However, this regime is explored
in the set of free shear layers experiments of \citet{Bid92}, who
show that the flow remains turbulent\footnote{By virtue of the
Taylor-Proudman theorem, the flow should eventually become
bidimensional but this happens only at higher values of $|S|$.}
(although linearly stable) down to $S\sim -2$ for Reynolds
numbers\footnote{Notice that, as the turbulent shear layer widens
downstream, \citet{Bid92} base their definition of the Reynolds
number on the downstream distance $x$, which needs to be related
to the layer width from which all Reynolds numbers quoted here are
defined, and which is referred to as $2\delta_M$ in their paper.
The two quantities can be related with the help of the various
relations given in section 3 of their paper. This amounts to
reducing the Reynolds numbers they quote by a factor $\sim 7$.
Finally, the number given above corresponds to the most downstream
point of measurement, where the flow should be closest to a
developed (rather than developing) turbulent flow (incidentally,
this is much farther downstream than the region where the pictures
shown in the paper are taken).} $\sim 4000$ (see figures 14 and 16
of their paper). On the other hand, in the numerical simulations
of anticyclonic ($S<0$) rotating Couette flows of \citet{Bech97}
($Re\sim 5000$) and \citet{Komm96} ($Re\sim 3000$), turbulence is
lost\footnote{Rotation in these numerical experiments is
characterized by a global rotation number $R_o\equiv 2\Omega h/U$
rather than by the local rotation parameter $S$. In the central
part of the profile, one usually has $R_o\gtrsim 0.2 |S|$ for
fully turbulent flows, but it is difficult to precisely relate the
relative level of rotation in these experiments to the critical
limit between linearly stable and unstable rotating flows.} for
$S\sim -1$ in the central part of the flow. This situation is
similar to the one relating the simulations of \citet{BHS96} and
\citet{HBW99} to the experimental data of \citet{Tay36} and
\citet{Wen33} quoted in \citet{RZ99}; this analogy will be further
discussed in section~\ref{ssturb} and \ref{cor}.

\subsection{Couette-Taylor flows}\label{CTF}

\begin{figure}
 \plotone{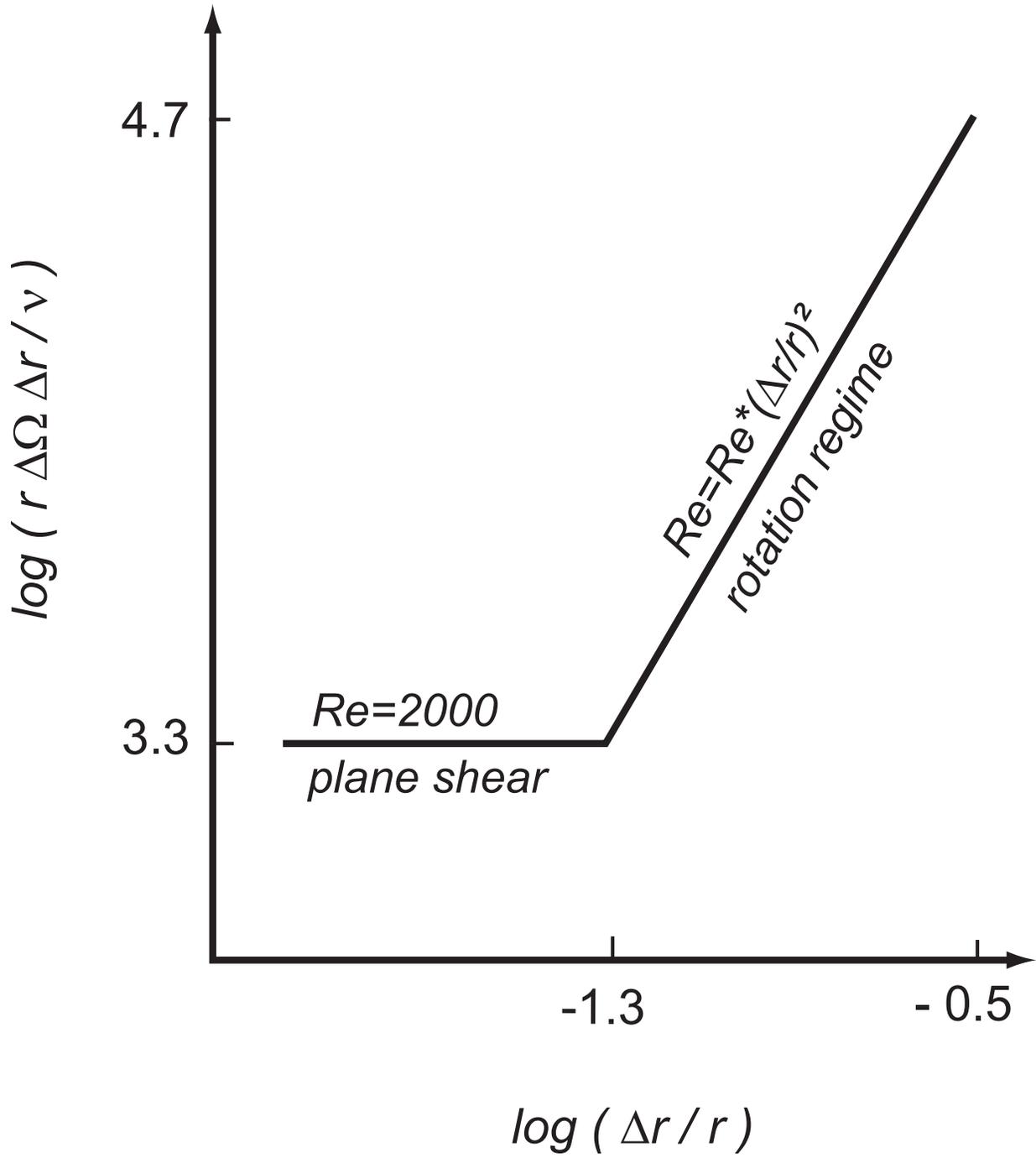}
  \caption{\small Idealized behavior of the minimal Reynolds number of fully
turbulent Couette-Taylor flows, as a function of the relative gap
width.}\label{fig3}
\end{figure}

Couette-Taylor flows are produced from two concentric rotating
cylinders. Most investigations of this type have focused on the
linearly unstable regime. The linearly stable one, which is more
directly relevant to astrophysics, has only been explored by
\citet{Tay36}, who maintained the inner cylinder at rest, and by
\citet{Wen33}, who also reported results when the flow is close to
marginal stability (i.e. linearly stable, but close to constant
specific angular momentum). The Reynolds number of these flows is
defined as

\begin{equation}\label{ReCT}
  Re= \frac{r \Delta \Omega \Delta r}{\nu}
\end{equation}

\noindent where $r$ is the mean of the two cylinder radii, and
$\Delta\Omega$ and $\Delta r$ are respectively the difference in
angular velocity and the gap width of the two cylinders. Both
investigations mentioned above did characterize the behavior of
the minimal Reynolds number for well-developed turbulence to be
maintained as a function of the cylinders relative gap width; this
behavior is sketched on Fig.~\ref{fig3}. Recently, a French team
has undertaken an experimental investigation of flow profiles
which are approximately keplerian in the mean, and found that
turbulence was also maintained for Reynolds numbers of the order
of a few thousand for a relative gap width of the order of a third
\citep{Rich01}.

\noindent The minimal Reynolds number appearing in Fig.~\ref{fig3}
is obtained by starting from an initially laminar flow, and
progressively increasing the difference in angular velocity of the
two cylinders (or only the outer cylinder angular velocity if the
inner one is at rest). When starting from an initially turbulent
flow and reversing the process, the loss of turbulence occurs for
Reynolds numbers which can be significantly lower, but the flow is
then highly intermittent; it is reasonable to assume that the
minimal Reynolds numbers of Fig.~\ref{fig3} are characteristic,
albeit overestimated, values for well-developed turbulence
\citep{Rich01}.

The two remarkable features of this minimal Reynolds number are a
behavior which is similar to plane Couette flows for $\Delta
r/r\lesssim 1/20$, with $Re\simeq 2000$, and a quadratic scaling
[$Re\simeq Re^*(\Delta r/r)^2$ with $Re^*\simeq 6\times 10^5$]
which is characteristic of rotation, as argued by \citet{RZ99};
these authors also show that in the same regime, the turbulent
viscosity $\nu_t\simeq \beta r^3 |d\Omega/dr|$ with $\beta\simeq
10^{-5}$. A heuristic explanation of these features is presented
in section \ref{turbvis}.

The Navier-Stokes equation for these flows is most meaningfully
compared to that of other flows when substracting out the mean
flow rotation $\Omega_0$ (i.e., the average angular velocity of
the two cylinders), as only differential rotation plays a role in
the generation of turbulence. Defining ${\bf w}={\bf v}-\Omega_0 r
{\bf e}_{\phi}$, and $\phi=\theta-\Omega_0 t$ (so that ${\bf w}$
and $\phi$ are the velocity and azimuthal coordinate in the
rotating frame, respectively), the Navier-Stokes equation for
${\bf w}=(w_r,\ w_\phi,\ w_z)$ becomes

\begin{equation}\label{NSCT}
  \frac{\partial{\bf w}}{\partial t}+{\bf w}.\nabla'{\bf w}
  +2{\bf \Omega}\times{\bf w}-\frac{w_{\phi}^2}{r}{\bf e}_r
  +\frac{2w_{\phi}w_r}{r}{\bf e}_{\phi}=
  \left(-\frac{\nabla P}{\rho}+r\Omega^2{\bf e}_r\right) +
  \nu\Delta{\bf w},
\end{equation}

\noindent where ${\bf w}.\nabla'{\bf w}\equiv({\bf w}.\nabla w_r)
{\bf e}_r+({\bf w}.\nabla w_{\phi}) {\bf e}_{\phi}+({\bf w}.\nabla
w_z) {\bf e}_z$. For future reference, I refer to the terms
$w_{\theta}^2/r$ and $2w_rw_\theta/r$ as ``geometric terms", as
they arise because of the cylindrical geometry\footnote{Such terms
also arise in principle from the viscous term, but they are
inessential to the argument developed in this paper.}.

For these flows, the rotation parameter defined in Eq.~(\ref{Ro})
reads

\begin{equation}\label{RoCT}
  S=\frac{2\Omega}{rd\Omega/dr}=-\frac{2}{q},
\end{equation}

\noindent where $q\equiv -(r/\Omega)(d\Omega/dr)$ is the parameter
defined by \citet{BHS96} to characterize rotation profiles. The
flow is stable according to Rayleigh's criterion when $q<2$, i.e.
when $S<-1$, quite similarly to rotating Couette and free flows,
although the processes through which instability occurs are
different. Note also that Eqs.~(\ref{NSR}) and (\ref{NSCT}) differ
only through the geometric and centrifugal terms.

The fact that the minimum Reynolds number for developed turbulence
is identical in plane Couette and Couette-Taylor flows with
$\Delta r/r\lesssim 1/20$ and the inner cylinder at rest can be
understood in the following way. First, the advection term (which
is the source of the turbulence cascade as indicated by the very
existence of the Reynolds number) dominates over the geometric
terms when $r\Delta\Omega/\Delta r\gg r\Delta\Omega/r$, i.e.
$\Delta r/r\ll 1$. Second, $\Delta\Omega=\Omega$ (one cylinder
being at rest), so that the Coriolis term is also very small
compared to the advection term, and Eq.~(\ref{NSCT}) nearly
reduces to Eq.~(\ref{NSC}). Note furthermore that the Coriolis
force does not appear to significantly affect the minimal Reynolds
number for the onset of turbulence for the values of $q$ of
interest here (i.e., $q\gg 1$ to $q\sim 1$ - $2$), both in the
limiting plane Couette regime and in the rotation regime, as
exemplified by the data of \citet{Wen33} for nearly neutral flows,
which follow the same law for the minimal Reynolds number, down to
the plane Couette limit.

\subsection{Shearing sheet}\label{SHESHE}

Accretion disk flows in the shearing sheet approximation are
closely related to the Couette-Taylor flows previously described.
They differ in only three respects.

First, the mean angular velocity profile $\langle\Omega\rangle$ in
Couette-Taylor flows is imposed by the boundary conditions, and by
the condition of stationarity of the mean viscous or turbulent
transport of angular momentum, from the azimuthal momentum
equation (with the walls acting as source and sink of angular
momentum). The radial momentum equation then imposes the mean
radial pressure profile $\langle P\rangle$, and the resulting
tidal force term $(-\nabla \langle P\rangle/\langle\rho\rangle
+r\langle\Omega\rangle^2)$. On the contrary, in accretion disks,
the mean radial angular velocity profile mostly results from the
gravitational attraction of the central body, which imposes a
nearly keplerian profile in cold disks, but the disk is never
globally stationary, due to viscous/turbulent transport
(nevertheless, an approximate stationarity is nearly achieved
locally on the dynamical timescales of interest for the onset of
turbulence). Therefore, in keplerian disks in the shearing sheet
approximation, the tidal force term $(-g+r\langle\Omega\rangle^2)$
is the source of the (keplerian) angular velocity profile and not
its consequence. Furthermore, one usually neglects the radial
pressure gradient locally, and assumes that the gravitational
force has cylindrical (and not spherical) symmetry for simplicity,
as cold disks are thin.

Secondly, a local approximation is performed, by restricting
consideration to a radial box of width $\Delta r\ll r$; one also
usually assumes that the height of the box is comparable to its
width. Under these assumptions, one neglects the geometric terms
in Eq.~(\ref{NSCT}), and describes the flow in local cartesian
coordinates ($x\leftrightarrow r$, $y \leftrightarrow r\phi$ where
$\phi$ is the azimuthal coordinate in the rotating frame
introduced for Couette-Taylor flows). One also linearizes the
angular velocity profile.

Finally, this local approximation and the resulting change of
geometry from cylindrical to cartesian (except for the Coriolis
force term which is kept) allows one, in numerical simulations, to
adopt a particular form of the periodic boundary condition in the
radial direction, in which the fluid quantities on the radial
boundaries are longitudinally displaced all the time with the mean
angular velocity difference during a time step before the periodic
boundary condition is applied (see \citealt{HGB95} for details on
this procedure).

With these prescriptions (aside from the boundary conditions), the
Navier-Stokes equation, in the shearing sheet approximation and in
the rotating frame, becomes

\begin{equation}\label{NSSS}
\frac{\partial{\bf w}}{\partial t}+{\bf w}.\nabla{\bf w}
  + 2{\bf \Omega}\times{\bf w} =
  - \frac{\nabla P}{\rho} +
    2q\Omega^2x{\bf e}_r + \nu\Delta{\bf w},
\end{equation}

\noindent where $x=r-r_0$ and $r_0$ is the position of the center
of the shearing sheet box. The term $2q\Omega^2x$ represents the
tidal force (difference of the gravitational and inertial force);
$q$ is the parameter introduced in the previous subsection and
measures the steepness of the rotation profile. Note that the
pressure term contains only fluctuations related to the presence
of turbulence, which is not the case in Couette-Taylor flows. It
is interesting to note that this equation shares features with
both Eqs.~(\ref{NSR}) and (\ref{NSCT}); in particular, linear
stability is ensured for $q<2$, i.e. $S<-1$ for the laminar linear
profile. This makes the loss of turbulence in the simulations of
\citet{BHS96} and \citet{HBW99}, for values of $q$ smaller than 2
by a few percents only, all the more intriguing.

\subsection{Shearing sheet, rotating Couette flows, and
turbulence}\label{ssturb}

In fact, all available pieces of evidence strongly suggest that
numerical simulations of rotating Couette flows and of tidally
driven sheared motions in the shearing sheet limit should display
turbulence, as I argue now.

First, plane Couette flows, rotating Couette flows, and tidally
driven shearing sheet flows have similar linear stability
properties. For all three types of flows, the viscously relaxed
laminar solution is a simple linear shear, which is always
linearly stable for the plane Couette flow\footnote{I consider
unbounded flows in this discussion, as instabilities due to the
boundary in viscous fluids are not relevant in astrophysics.}, and
stable for the other two flows once $S< -1$ (which is the only
case of interest here). Plane Couette flows are subject to finite
amplitude instabilities (see, e.g., \citealt{LK88},
\citealt{DZ91}, and references therein). The same is true of
rotating Couette flows \citep{John63}, and of shearing sheet flows
\citep{Bulle93}. As finite amplitude instabilities are considered
to trigger the turbulence seen both in experimental and numerical
investigations of plane Couette flow, one would expect the same to
be true of the other two flows.

Secondly, let us reexamine the differences between rotating
Couette flows and the shearing sheet flows with the other flows
discussed previously. They amount to differences in boundary
conditions, of mean force terms, and of geometry.

The shearing sheet boundary conditions are in a way intermediate
between rigid and free boundary conditions, as they imply that the
mean flow obeys rigid boundary conditions, whereas the fluctuating
part obeys periodic boundary conditions; rotating Couette flow
simulations are usually performed with rigid boundary conditions.
On the one hand Couette-Taylor flows implement rigid boundary
conditions. Although in real experiments, the vibrations of the
boundary play some role in triggering turbulent motions, there is
little doubt that in these experiments, turbulence is
self-sustained. On the other hand, from the experiments of
\citet{Bid92} the rotating shear flows with free boundary
conditions are turbulent, although by construction no mean steady
state can be reached in these systems. Therefore, it seems
unlikely that boundary conditions play an important role in the
presence or absence of turbulence in numerical experiments.

In the shearing sheet approximation, the mean shear is imposed by
the tidal force term; in rotating Couette simulations, it results
from the boundary conditions. In Couette-Taylor flows, the
boundary conditions do not only produce the shear, but also
generate a mean radial pressure gradient. Note however that the
term $-d\langle P \rangle/dr/\langle\rho\rangle
+r\langle\Omega\rangle^2$ of Eq.~(\ref{NSCT}) is similar in
function to the term $2q\Omega^2 x$ in Eq.~(\ref{NSSS}).
Furthermore, the mean pressure gradient in Couette-Taylor
experiments is radial, whereas it is longitudinal (streamwise) in
the rotating free shear layer experiments of \citet{Bid92}. This
suggests that neither large scale mean pressure gradients, nor
tidal terms, make any significant difference on the question of
the onset of turbulence in the various flows considered here,
especially that all gradient terms get out of the way in
incompressible flows (they disappear from the vorticity equation).

Finally, I will show in the next section that the main effect of
the geometry (which enters through the geometric terms in
Couette-Taylor flows) is to change the conditions of onset of
turbulence, but this does not affect the occurrence of turbulence
in itself.

Although such arguments do not exclude more complex possibilities
(as, e.g., that turbulence might be impeded in shearing sheet
flows by a combination of these factors instead of only one of
them), this strongly indicates that rotating Couette flows and
shearing sheet ones should be turbulent, suggesting that the
absence of turbulence in all the published simulations of this
kind stems from limitations in the numerics involved. This last
point is addressed in the next section.

\section{Phenomenology of subcritical turbulence}\label{turbvis}

The purpose of this section is to point out important features of
turbulence in sheared flows, through a phenomenological model
developed in section \ref{turbord}. The consequences of this model
are used in section \ref{cor} to identify the potential
limitations in the numerics just mentioned.

\subsection{Turbulent viscosity and the Kolmogorov
prescription}\label{turbvis1}

In a picture where the fluctuating turbulent scales can be
separated from the more regular large ones, it is meaningful to
write down an equation for both the mean $\langle X \rangle$ and
fluctuating $\delta X$ parts of any quantity $X$. In particular,
the evolution of the mean velocity reads

\begin{equation}\label{NSav}
  \frac{\partial \langle {\bf v}\rangle}{\partial
  t}+\nabla\cdot\langle\overline{\overline{\delta{\bf v }\delta{\bf v}}}\rangle=
  -\frac{\nabla \langle P \rangle}{\rho}+\nu\Delta\langle {\bf v}\rangle,
\end{equation}

\noindent where possible geometric and/or inertial terms have been
omitted for simplicity, as well as the effect of compressibility.
In the simple shear configurations of interest here, only the
$\langle{\delta v_y \delta v_x}\rangle$ (or $\langle\delta v_r
\delta v_{\phi}\rangle$) part of the Reynolds stress tensor is
relevant for radial turbulent transport.

By describing turbulent fluctuations with a characteristic
coherence scale $l_M$ and velocity amplitude $v_M$, \citet{Pr25}
argued that

\begin{equation}\label{reynolds}
  \langle\delta v_y \delta v_x\rangle\sim v_M^2
  \sim \nu_t\frac{d\langle v_x\rangle}{dy},
\end{equation}

\noindent with

\begin{equation}\label{nut}
  \nu_t\sim l_M v_M.
\end{equation}

\noindent Note that in cylindrical geometry $\langle\delta v_r
\delta v_{\phi}\rangle \sim\nu_t rd\langle\Omega \rangle/dr$.

The reasoning behind this formulation is similar to the one
relating the usual molecular viscosity to the molecular mean free
path and velocity dispersion (i.e., turbulent transport occurs
over a ``mean free path" $l_M$ with ``velocity dispersion" $v_M$);
Eq.~(\ref{reynolds}) can also be derived from more rigorous
multi-scale expansion techniques. In a Kolmogorov cascade picture,
$l_M$ is the energy injection scale (and characterizes the
coherence length of the largest eddies of the cascade), and $v_M$
the amplitude of the velocity fluctuations at this scale, as the
velocity amplitude decreases with decreasing scale in a Kolmogorov
spectrum. However, this does not mean that larger fluctuating
scales are not present in the flow, nor that they have no
influence in the development of turbulence; it just implies that
they dominate neither the energy spectrum nor the turbulent
transport.

An important feature of the turbulent viscosity prescription is
that the rate of energy dissipation $\epsilon$ --- which is also
the rate of energy transfer in eddy-scale (Fourier) space --- is
simply given by

\begin{equation}\label{energy}
  \epsilon\sim v_M^3/l_M\sim \begin{cases}
  {\nu}_t\left(d\langle v_x\rangle/dy\right)^2 & \text{(cartesian),}\\
  \nu_t\left(r d\langle\Omega\rangle/dr\right)^2 & \text{(cylindrical),}
  \end{cases}
\end{equation}

\noindent as can be shown most directly by deriving the relevant
macroscopic energy equation. Eqs.~(\ref{reynolds}) and
(\ref{energy}) imply in particular that the characteristic
frequency of turbulent motions is the shear frequency, i.e.

\begin{equation}\label{freqcar}
  \frac{v_M}{l_M}\sim \begin{cases}
   d\langle v_x\rangle/dy & \text{(cartesian),}\\
   r d\langle\Omega\rangle/dr & \text{(cylindrical).}
  \end{cases}
\end{equation}

This reflects the fact that an externally imposed shear locally
possesses no characteristic scale (besides the scale of the flow),
but only a characteristic frequency, so that shear turbulence can
only couple efficiently to the shear if its characteristic
frequency or coherence time (at the energy injection scale imposed
by the mechanism which drives turbulence) matches the shear
frequency\footnote{As the coherence time of smaller scale eddies
is shorter, they are less or little affected by the shear. As a
consequence, in a first approximation, the turbulence is more or
less isotropic at scales $< l_M$, and anisotropy is ignored in the
whole argument.}.

The turbulent viscosity description has been applied to a wide
variety of setups to describe the mean properties of turbulent
flows, both in the the vicinity of walls and in the main part of
either free or bounded flows (e.g., \citealt{Tenlum};
\citealt{les}).

\subsection{Turbulence scales : phenomenological model and orders of magnitude}\label{turbord}

My primary purpose here is to point out some interesting
consequences of the turbulent viscosity prescription. Indeed, one
expects that a flow undergoes a transition to turbulence when the
turbulent transport becomes more efficient than the laminar one
for subcritical flows. This implies that

\begin{equation}\label{nutlim}
  \nu_t\gtrsim \nu\ \text{when $Re\gtrsim {Re}_m$,}
\end{equation}

\noindent where ${Re}_m$ stands for the minimum Reynolds numbers
for the onset of turbulence discussed in the previous section. For
example, note that for Couette-Taylor flows, from the data of
\citet{Wen33} and \citet{Tay36} $\nu_t/\nu\sim\beta Re^*\sim 6$,
where $\beta$ and $Re^*$ are the quantities introduced in the
previous section in the discussion of these flows; furthermore,
when the minimal Reynolds number is searched for by decreasing the
velocity difference between the cylinder from an initially
turbulent state instead of increasing it from an initially laminar
one, the ratio $\nu_t/\nu$ is sensibly closer to unity
\citep{Rich01}.

Away from boundary layers (if any), the only scales which are
relevant for characterizing the shear gradient are the typical
size of the shear flow, $\Delta y$ (resp. $\Delta r$) in cartesian
(resp. cylindrical) geometry, and the typical shear amplitude over
this scale, $\Delta v_x$ (resp. $r\Delta\Omega$). Combining Eqs.
(\ref{reynolds}), (\ref{energy}) and (\ref{nutlim}) then yields,
for the bulk of the turbulent flow

\begin{equation}\label{lm}
   l_M\sim \begin{cases}
   \Delta y/{Re}_m^{1/2} & \text{(cartesian),}\\
   \Delta r/{Re}_m^{1/2} & \text{(cylindrical),}
   \end{cases}
\end{equation}

\noindent and

\begin{equation}\label{vm}
   v_M\sim \begin{cases}
   \Delta v_x/{Re}_m^{1/2} & \text{(cartesian),}\\
    r\Delta\Omega/{Re}_m^{1/2} & \text{(cylindrical).}
   \end{cases}
\end{equation}

I wish to stress that Eqs.~(\ref{lm}) and (\ref{vm}) do not imply
that turbulence is a global rather than local phenomenon. On the
contrary, Eqs.~(\ref{energy}) and (\ref{freqcar}) relate $l_M$ and
$v_M$ to local characteristics of the mean flow. Note that these
relations justify (at least for subcritical flows) the separation
of scales between the mean large scale flow and the fluctuating
small scale one which is assumed in the turbulent viscosity
description, because $Re_m$ usually exceeds a few thousands.

These relations have a direct physical interpretation. Consider
for example two planar Couette flows with identical shear rates,
and with wall spacing $\Delta y$ and relative velocity $\Delta
v_x$ which differ by a given ratio. Obviously, the scaling with
$\Delta y$ and $\Delta v_x$ is a natural consequence of the
scaling similarity between flows which are otherwise $\it
identical$. On the other hand, consider {\it different} flows,
with identical shear rates, but different minimal Reynolds numbers
(e.g., plane Couette and Couette-Taylor flows with appropriate
parameters). A larger minimal Reynolds number is a sign of a
greater difficulty to trigger turbulence, i.e. an increased
difficulty for turbulent transport to dominate over the viscous
one, and therefore is a sign of a smaller scale turbulence, due to
the physical picture underlying the turbulent viscosity
prescription (i.e., the transport occurs over a smaller ``mean
free path" $l_M$, and correlatively with a smaller ``random
velocity" $v_M$ due to the assumption of identical shear rate
between the two different flows).

From these relations, one can easily check that, at the minimum
Reynolds number, the advection term, which dominates scale
coupling and is the primary cause of the inertial turbulent
spectrum, is comparable to the dissipation term, at the turbulent
transport scale. As a consequence, the turbulence possesses little
or no inertial domain at its threshold. Furthermore, as long as
there is no change in the turbulence generating process,
increasing the Reynolds number can only result in lowering the
dissipation scale with respect to $l_M$, and therefore in the
progressive build up of an inertial spectrum (e.g., imagine one
does this by reducing the viscosity while maintaining the large
scale structure of the flow unchanged).

It is important to notice that the estimates of Eqs.~(\ref{lm})
and (\ref{vm}) remain valid for Reynolds numbers larger than the
turbulence threshold, as long as the turbulence generating process
is unchanged. The predictions of the scaling proposed here are
well supported by the available empirical and numerical evidence,
as shown in Appendix A.

\subsection{Consequences: Couette-Taylor flows}\label{CTFS}

Eqs.~(\ref{lm}) and (\ref{vm}) have particularly interesting
consequences for the understanding of turbulence in Couette-Taylor
flows. For definiteness, I will first focus on flows where the
inner cylinder is at rest. As argued at the end of section
\ref{CTF}, for $r\gg\Delta r$, the Navier-Stokes equation for
Couette-Taylor flows [Eq.~(\ref{NSCT})] then reduces to the
Navier-Stokes equation for planar Couette flows [Eq.~(\ref{NSC})]
and the minimal Reynolds number is constant. However, when $\Delta
r\rightarrow r$, the geometric terms $O(w^2/r)\sim
(r\Delta\Omega)^2/r$ become comparable to the advection one on
scale $\Delta r$. Furthermore, if, at some radial location $r$ in
the flow, $Re_m$ remained constant when $\Delta r\gg r$,
Eq.~(\ref{lm}) would imply that $l_M$ could become arbitrarily
larger than $r$, which makes little sense. In fact, one expects
that $l_M\propto r$ once $\Delta r/r$ exceeds some critical ratio
$\Delta_c$ (which for the time being is expected to be of order
unity), for two reasons: first, the geometric terms introduce a
limiting scale (the radius $r$), which must be accounted for by
the turbulent viscosity description\footnote{This reasoning is
somewhat similar to the one which imposes that $l_M\propto y$ in
the vicinity of the wall in Couette or channel flows, and which
has lead to the derivation of the well-known ``law of the wall",
describing the mean structure of turbulent flows close to the wall
(e.g., \citealt{LL,Tenlum,les}).}; second this prescription for
$l_M$ is necessary to satisfy the requirement that $w\nabla
w\gtrsim w^2/r$ at the largest scale of the inertial spectrum (in
order to maintain such a spectrum). Consequently, let us assume
that

\begin{equation}\label{lmr}
  l_M\sim \gamma r,
\end{equation}

\noindent when $\Delta r/r > \Delta_c$, and where $\gamma$ is a
constant to be determined later. The argument presented here
suggests that $\Delta_c\sim 1$ whereas the data imply that it is
significantly smaller than unity (see below); as for the large
values of the minimal Reynolds numbers for turbulence (which one
would also naively expect to be of order unity), this originates
in the (still unknown) mechanism which sustains turbulence.

Eqs.~(\ref{lm}) and (\ref{lmr}) must be satisfied simultaneously,
and this is possible only if $Re_m$ depends on the relative gap
width:

\begin{equation}\label{RemCT}
  Re_m\sim\frac{1}{\gamma^2}\left(\frac{\Delta r}{r}\right)^2,
\end{equation}

\noindent which explains the behavior seen on Fig.~(\ref{fig3}).
Equivalently\footnote{Remember that $\Delta r$ and $\Delta\Omega$
have been introduced in Eqs.~(\ref{lm}) and (\ref{vm}) to
represent local gradients in order of magnitude.}, $r^3(d\Omega/d
r)/\nu \gtrsim 1/\gamma^2$; this shows that, as soon as $\Delta
r\gtrsim \Delta_c$, the width of the flow does not influence the
onset of turbulence, which becomes a purely local phenomenon.

\noindent The velocity fluctuation amplitude now reads

\begin{equation}\label{vmr}
  v_M\sim\gamma\frac{r^2\Delta\Omega}{\Delta r}\simeq\gamma r^2\frac{d\Omega}{dr},
\end{equation}

\noindent i.e., it is proportional to the local shear rate. As a
consequence, the turbulent viscosity becomes

\begin{equation}\label{nutr}
  \nu_t\sim\gamma^2 r^3\frac{d\Omega}{dr}.
\end{equation}

\noindent A similar relation has also been proposed by
\citet{RZ99} directly from experimental torque data. Note that the
reasoning leading to Eq.~(\ref{nutr}) implicitly assumes that the
flow compression plays little role, so that this result may not
necessarily apply to supersonic turbulence. In order for
Eqs.~(\ref{RemCT}) and (\ref{nutr}) to faithfully account for the
properties\footnote{Most notably, the scaling of the Reynolds
number with $(\Delta r/r)^2$ in the rotation regime, and the near
coincidence between $\beta$ and ${Re^*}^{-1}$.} of Couette-Taylor
flows described in section \ref{CTF}, one needs to tie up a few
loose ends:

\begin{itemize}
  \item Because the Coriolis force does not seem to affect the minimal
  Reynolds number of turbulence (see the closing comment of section \ref{CTF}),
  the argument above must apply to any value of the $q<2$
  (the parameter introduced in section \ref{SHESHE} to
  characterize the local rotation profile), and not only to situations
  with the inner cylinder at rest; however, for $q\sim 1$, the geometric term is
  always comparable to the advection term, and the argument is less transparent.

  \item The gap relative width in Fig.~\ref{fig3} is measured with respect
  to the mean radius of the rotating cylinders, whereas a local value is used above.
  However, the relative gap widths shown in this figure are all
  sufficiently smaller than unity to make the difference between
  the two quantities negligible in the scaling argument developed
  here. Incidentally, this shows again that turbulent properties
  are local; e.g., turbulent eddies become larger when one
  moves outwards in a sufficiently wide cylindrical system.
  Correlatively, Eq.~(\ref{lmr}) follows also directly from the
  fact that $r$ is the only available local scale.

  \item The relations derived above imply that
  $\gamma^2=\beta={Re^*}^{-1}$, but the values quoted in section
  \ref{CTF} for the last two quantities differ by a factor of 6.
  However, it was also pointed out there that the value of $Re^*\simeq 6\times 10^{5}$
  is overestimated, because it leads to a ratio $\nu_t/\nu$ which
  is too large due to the particular experimental protocol
  adopted by \citet{Tay36} and \citet{Wen33}. Also, recent (still
  unpublished) experiments on Couette-Taylor flow in the
  ``keplerian" regime ($q=3/2$) exhibit sustained turbulence for
  Reynolds numbers smaller than the limit of Fig.~2, but with a
  different experimental protocol \citep{Rich01}.
  Therefore, it is reasonable to
  assume that $\beta$ is a much better measure of $\gamma^2$ than
  $Re^*$; this assumption is made in the remainder of this paper, and $\beta$
  is used everywhere instead of $\gamma^2$.
\end{itemize}

The critical value of the relative gap width $\Delta_c$ which
separates the planar regime from the rotating one obtains when the
values of $l_M$ in both regimes are equal. This yields

\begin{equation}\label{Deltac}
 \Delta_c\equiv \left(\frac{\Delta r}{r}\right)_c\sim \left(\beta Re_p\right)^{1/2},
\end{equation}

\noindent where $Re_p\sim 2000$ is the minimal Reynolds number in
the planar limit. This gives $\Delta_c\sim 1/7$, which is somewhat
larger than the value of $1/20$ shown on Fig.~\ref{fig3} (but
closer to the uneducated guess $\Delta_c\sim 1$), because of the
reduction adopted above of the value of $Re^*$.

Note also that $l_M\sim r/300$. One might wonder why such a small
length scale arises, whereas one would naively expect $l_M\sim r$
on dimensional grounds. However, the same dimensional type of
argument would also predict that turbulence sets in for $Re\gtrsim
1$, which is strongly violated by the empirical evidence. The two
facts have the same physical origin: the (as yet not understood)
mechanism which sustains turbulence.

Two other explanations of the behavior of the Reynolds number with
relative gap width have previously been proposed in the
literature. \citet{Zel81} assumed that turbulence in these
Couette-Taylor flows is controlled by a competition between the
epicyclic (stabilizing) frequency and the shear rate which is the
source of the turbulent motions; however his findings are
inconsistent with some of the data (see the discussion of this
point in the appendix of \citealt{RZ99}). \citet{Bulle93} looks
for an explanation in terms of finite amplitude instabilities in
the WKB approximation, but this is incompatible with the fact that
the scale $r$ plays a key role in the problem.

I conclude this section by pointing out that the Coriolis force
appears nowhere in the arguments presented in this section, which
suggests that it plays little role in the development of
turbulence in subcritical Couette-Taylor flows, at least for
$q\sim 1-2$. Indeed, in opposition to the inertial (geometric)
terms, the Coriolis force does not single out any length scale. In
particular, the ratio of the advection term ($\sim w\nabla w$) to
the Coriolis one ($\sim w\Omega$) in Eq.~(\ref{NSCT}) is $\sim 1$
both at scale $l_M$ and at scale $\Delta r$ for the values
$|q|\sim 1-2$ of interest here, and increases with decreasing
scale in a Kolmogorov cascade picture. However, it does play a
role in the loss of turbulence in simulated rotating flows, but
this apparent paradox cannot be investigated in the framework of
the order of magnitude arguments developed in this section. The
next section is devoted to a discussion of this point.

\subsection{The role of the Coriolis force: beyond orders of
magnitude}\label{cor}

The question I want to address here is the following : why is
turbulence lost in numerical simulations of sheared flows when
even a small amount of rotation is added (and the resulting flow
remaining linearly stable) --- in particular for Couette flows and
disk flows in the shearing sheet approximation --- whereas in
experiments as different as the Couette-Taylor and rotating free
shear layers, it is maintained (in the same conditions of linear
stability). The forms of the Navier-Stokes equation for these
flows given in section \ref{phen} strongly suggests that this is a
consequence of the Coriolis force, as this is the only new force
term which is taken into account when rotation is added to free
shear layers and planar Couette flows in these simulations.

More specifically, clues to the role of the Coriolis force can be
found by inspecting the behavior of plane Couette and free shear
flows with and without rotation. The most noticeable and important
feature is that, in numerical simulations of rotating Couette
flows, even a small Coriolis term is able to suppress the very
largest scales of the turbulent motions. This is particularly
obvious, e.g., when comparing Figs.~7 and 20 of \citet{Komm96},
which shows that the very large scales which develop in the
streamwise direction in turbulent flows break up for rotation
numbers as small as a few percents. This feature is quite
understandable on the basis of the velocity spectra shown in
\citet{Bech95}, which imply that $kv_k$ is most likely sensibly
smaller than $\Omega$ at scales larger than $l_M$. A similar
feature can also be indirectly found in \citet{Bid92} (see their
Fig.~11 and 14, to be combined with their Fig.~16), who show that
the Reynolds stress tensor magnitude decreases by a
factor\footnote{The noise in the data at large rotation number
does not permit a very precise estimate of this reduction factor,
but the value of 10 quoted here seems a bare minimum.} of at least
$10$ when the parameter $S$ introduced in Eq.~(\ref{Ro}) varies
from 0 to $\lesssim - 1$; this suggests that the size of the
largest turbulent scales in these flows is also substantially
reduced under the action of the Coriolis force\footnote{Note that
in \citet{Bid92}, as pointed out by the authors themselves,
turbulence is not lost, and the flow remains three-dimensional,
although the velocity fluctuations anisotropy is clearly affected
by rotation.}. This indicates that, although Eq.~(\ref{lm}) always
provides reliable orders of magnitude for $l_M$, it underestimates
the relevant eddy scale by a factor $\sim 3$ for non-rotating
flows (see Appendix A), while overestimating it by at least the
same factor once rotation is introduced. As a consequence the loss
of turbulence in the numerical simulations of rotating Couette
flows of \citet{Bech97} and \citet{Komm96} is clearly an effect of
the limited small scale resolution due to the large box sizes
(especially in the streamwise direction) adopted in these works:
the smallest available scales do not allow these authors to
account for the inertial part of the energy spectrum, while all
the larger scales are wiped out by the Coriolis force. It is more
than likely that, in these simulations, the Coriolis force kills
the large scale mechanism which has been identified to sustain
turbulence in plane Couette flow simulations (\citealt{Jim91};
\citealt{Wal97}; \citealt{Ham95}; see section \ref{plcou}). The
fact that free rotating layers and Couette-Taylor flows remain
turbulent at larger levels of rotation than the ones to which
turbulence is lost in these simulations implies that a different
mechanism for sustaining turbulence is at work in these flows, and
that it operates at scales comparable to, but apparently smaller
than, the estimate of Eq.~(\ref{lm}). This other mechanism has not
yet been found in numerical simulations. It would be interesting
to know whether this change of mechanism is related to the fact
that the Coriolis force apparently selects the direction of
instability of finite amplitude defects \citep{John63}.

The same line of argument applies to the shearing sheet
simulations of \citet{BHS96} and \citet{HBW99}. Indeed, the
effective Reynolds number of these simulations is not an issue, as
the code used by \citet{HGB95} and \citet{HBW99} is able to find
turbulence --- or at least the large scale mechanism already
alluded to --- in non-rotating Couette flows, and this happens
only for Reynolds numbers larger than at least $1500$. Also, the
argument developed in section \ref{turbord} shows that the
Coriolis force by itself should not change the minimal Reynolds
number for the onset of turbulence, an inference confirmed by the
fact that turbulence is seen developing in the rotating free shear
layer of experiments of \citet{Bid92} for roughly comparable
Reynolds numbers. Under the assumption (cf the arguments developed
above) that Eq.~(\ref{lm}) provides an estimate for the largest
turbulent scale which is overestimated by a factor of at least $3$
in the presence of a Coriolis force term, one obtains $l_M\lesssim
\Delta y/100$, with (possibly much) smaller values more than
likely. This is most probably too close to the largest resolution
achieved in the shearing sheet simulations (the number of zones in
any direction being no larger than 250), especially when
artificial viscosity is taken into account, for turbulence to show
up in these simulations.

To conclude this section, it is worth noting that some other
numerical questions must be considered to find turbulence in these
systems. First, it is well known from experiments with subcritical
flows that the way perturbations of the flow are designed has an
influence on the appearance of turbulence. This suggests that some
care must be exercised in the choice of the initial conditions in
numerical experiments; in particular, it might be useful to ensure
that at least some condition of finite amplitude instability is
satisfied in this choice. Secondly, the role of the choice of the
Courant number is not completely obvious, even in situations where
the CFL condition is not violated. For example, in a series of yet
unpublished simulations of linearly stable Couette-Taylor flows
performed with the Zeus code in collaboration with David Clarke,
we did initially find that turbulence would set in for flows which
are ``not too far" from planar Couette flows (including some
roughly keplerian flows), but it would eventually disappear when
reducing the maximal allowed time-step, although the CFL condition
was satisfied in all runs. The reason of this behavior is not yet
completely elucidated, but it appears to have some direct
connection to the question of resolution just discussed\footnote{A
counterintuitive dependence of hydrodynamic simulations on the
Courant number is also visible on Fig.~1 of \citet{PW94}.}. In any
case, the disappearance of turbulence in numerical simulations of
Couette-Taylor flows which are experimentally known to be
turbulent is a serious cause of worry on the reliability of the
conclusions drawn from the published shearing sheet numerical
experiments.

\section{Summary and astrophysical implications} \label{dis}

Eqs.~(\ref{lm}) and (\ref{vm}) (section \ref{turbord}) along with
their consequences constitute the central findings of this paper.
They express the natural length and velocity scales which are
involved in turbulent transport in subcritical flows in terms of
the local mean characteristics of the flow, and result from the
constraint that the turbulent transport dominates over the viscous
one in the framework of the turbulent viscosity description. The
scaling and orders of magnitudes implied by these relations is
supported by the available experimental and numerical evidence
(see the appendix and the beginning of section \ref{cor}).

These scaling relations have two important consequences. First
they provide an explanation for the minimum Reynolds number
dependence on the relative gap width in Couette-Taylor
experiments, displayed on Fig.~\ref{fig3} (section \ref{CTFS}). A
theoretical explanation of this behavior has long been sought for,
but none has satisfyingly been proposed yet; the phenomenological
one presented here has the advantage of connecting apparently
unrelated features, to be consistent with all the experimental
constraints, and to point out the direction in which such a
theoretical explanation might be looked for. Incidentally, the
existence of this phenomenological explanations strengthens the
validity of these scaling laws. Secondly, the comparison of the
various flows presented in section \ref{phen} implies that disk
flows described in the framework of the shearing sheet
approximation should be turbulent (section \ref{ssturb}), and the
scaling relations strongly suggest that the absence of turbulence
in the available shearing sheet numerical simulations is due to a
lack of resolution (section \ref{cor}). This follows because the
Coriolis force destroys large scale fluctuations, thereby
affecting in a major way the nonlinear mechanism through which
turbulence is maintained. At present, this mechanism is not
understood, except, to some extent, for plane Couette flows.

Understanding to which extent these results are helpful in
characterizing and quantifying turbulent transport in accretion
disks is an important issue. Three factors at least must be
accounted for: the magnitude of the disk pressure, the vertical
scale height, and the presence of a magnetic field; these factors
are not macroscopically independent, but relate differently to the
onset of turbulence.

The disk pressure affects the problem in two a priori different
ways: first, the turbulent transport picture presented in this
paper requires the underlying turbulence to be subsonic (see also
\citealt{HUR01}); second, turbulent velocity fluctuations require
a force to produce them, and only the pressure force is available
to this purpose in the hydrodynamical case, independently of the
details of the underlying mechanism which sustains this
turbulence. The first constraint is easily quantified: turbulent
motions are subsonic if $v_M/c_s \lesssim 1$ ($c_s$ is the sound
speed); in accretion disks, $c_s\sim \Omega H$, and from
Eq.~(\ref{freqcar}), this implies that $H \gtrsim l_M$ ($H$ is the
disk scale height). To quantify the second constraint, note that a
given fluctuating blob of size $l_M$ undergoes a velocity change
$\delta u\sim l_M (r d\Omega/dr)\sim l_M \Omega$ over a time-scale
$\sim v_M/l_M\sim \Omega$, because the coupling to the shear is
the source of turbulent motions at the largest scales; the largest
pressure variation at any scale is $\delta P/\rho\sim c_s^2$, and
requiring that the resulting pressure force at scale $l_M$ is able
to account for the turbulent velocity fluctuations at this scale
requires\footnote{This argument ignores the possibility of
supersonic turbulence.} again $H \gtrsim l_M$.

The turbulent scales ($\sim l_M$) are connected to the mean flow
scales through the mechanism which sustains turbulence. In an
accretion disk, only two such mean flow scales are available
locally: $H$ and $r$. The role of $r$ has already been discussed;
the role of the vertical scale height depends on the anisotropy of
the mechanism which sustains turbulence. In the absence of
constraint on the nature of this mechanism for rotating shear
flows, I will examine in turn two limiting assumptions:

\begin{itemize}
  \item This process is ``isotropic", i.e. the scales it requires
  to operate are roughly identical in all directions --- shearwise,
  streamwise and spanwise (this is the case for example of the
  nonlinear mechanism mentioned in section \ref{plcou} for non-rotating plane
  Couette flows). In this case, the elementary box in which this
  mechanism operates must be of size $H$, which implies in particular that
  $\Delta r \simeq H$ in all the relations used in the previous
  sections of this paper. If $H/r \lesssim \Delta_c$ [cf
  Eq.~(\ref{Deltac})], as expected in most disk models, the rotation
  regime of Couette-Taylor flows is irrelevant; instead, the shearing
  sheet approximation applies. As argued at the end of section
  \ref{CTF}, the Coriolis force is not expected to affect sensibly
  the minimal Reynolds number of turbulence, so that
   $\nu_t\sim (r\Delta\Omega)H/Re_p\sim 10^{-3}c_s H$, and the
  Shakura-Sunyaev parameter $\alpha\sim 10^{-3}$. Note in this
  case that the constraint $H\gtrsim l_M$ is always satisfied.

  \item The process is not sensitive to the vertical scale height
  except through the pressure requirement described above. As a
  consequence, as long as $H \gtrsim \beta^{1/2} r\sim 3.10^{-3}
  r$ (which is likely to be satisfied in accretion disks), the
  Couette-Taylor rotation regime applies, and $\nu_t\simeq
  \beta r^3 d\Omega/dr\sim \beta \Omega r^2$, so that the
  Shakura-Sunyaev parameter $\alpha\sim \beta(r/H)^2$ lies in the
  range $10^{-3}-10^{-1}$.
  If $H\lesssim \beta^{1/2} r$, the turbulence is supersonic. Note
  however that the extra energy dissipation taking place in shocks
  makes a supersonic turbulence more difficult to maintain, and the
  disk might heat up until $H \sim l_M$ is satisfied again, or $l_M$
  might decrease, i.e. the turbulence maintaining process might be
  affected and $Re_m$ increased, or the limiting case
  considered here does not apply\footnote{If the Reynolds
  number is large enough, the disk must be turbulent; this follows
  by considering a narrow enough disk portion so that $H$ exceeds
  its width, and at least one of the regimes of the previous sections
  does apply, inasmuch as boundary conditions are not essential to the
  onset of turbulence, as argued above}. This makes the relevance of
  supersonic turbulence to accretion disk theory unclear.
\end{itemize}

The conclusion of this brief discussion is that in the
Shakura-Sunyaev parameterization of the turbulent viscosity in
hydrodynamic disks, either $\alpha\sim 10^{-3}$, or $\alpha\sim
10^{-5}(r/H)^2$, depending on the unknown characteristics of the
mechanism which sustains turbulence. In principle, one should also
check that $Re > Re_m$; as $\nu\sim l c$ ($c\sim c_s$ is the
velocity dispersion and $l$ the mean free path), this translates
into $H/l \gtrsim 10^3$ in the first case above, and $r^2/Hl
\gtrsim \beta^{-1}\sim 10^5$ in the second, but both requirements
are most probably satisfied everywhere in astrophysical accretion
disks.

It is unclear how the presence of a magnetic field can modify
hydrodynamic shear turbulence. In particular, even a dynamically
non dominant field can easily affect the mechanism of generation
of turbulence, and therefore significantly modify the efficiency
of the turbulent viscosity transport, on top of adding a turbulent
resistivity, even if the MHD flow remains linearly
stable\footnote{This can happen, e.g., if the disk scale height is
small enough as not to let any magneto-rotational mode become
unstable, which is easily realized in disks with a near
equipartition between thermal and magnetic energies.}. Reversely,
the possible occurrence of hydrodynamic shear turbulence can
possibly affect in a major way our present understanding of MHD
transport and dynamo processes in accretion disks, which mostly
relies on the physics of the nonlinear development of the
magneto-rotational instability. Clarifying these questions is of
primary importance for accretion disk theory.

To conclude this paper, let me point out that there is one example
of keplerian disk which has been observed with a great luxury of
details, and which is not turbulent, namely Saturn's rings.
However, the requirements discussed above fail on several accounts
in ring systems, because both the particle size $d$ and mean three
path $l$ are comparable to $H$. For example, in the first limiting
case discussed above, the ring is necessarily laminar, while in
the second, because $H \gtrsim l_M$, the granularity of the system
makes scales $\lesssim l_M$ inaccessible to the fluid description;
the same argument makes supersonic turbulence most probably
irrelevant to ring systems.








\acknowledgments



\appendix

\section{Evidence for the proposed turbulent scaling}

Because ${Re}_m$ is at least of the order of $10^3$, the order of
magnitude estimates of Eqs.~(\ref{lm}) and (\ref{vm}) are sensibly
smaller than the mean flow length and velocity scales to which
$l_M$ and $v_M$ are usually assumed to be comparable.
Nevertheless, they are in good order of magnitude agreement with
the available evidence. Consider, for example, the simulation of
Couette flow reported in \citet{Bech95}, and further exploited in
\citet{Bech96b} to quantify the structure of the Reynolds stress
in the central region of Couette flows. For Couette flows Eq.
(\ref{vm}) gives $v_M\sim\Delta v_x/40$. The simulation just
mentioned has a Reynolds number\footnote{Note that our definition
of the Reynolds number differs from the one adopted in these
papers by a factor of 4.} of $5200$, i.e. well above the threshold
of transition to turbulence. The behavior of the Reynolds stress
as a function of the distance to the wall is represented on
Fig.~1a of \citet{Bech96b}, and, after accounting for the
particular normalization adopted in their graph\footnote{A
property of Couette flows is that the total mean shear stress
$\tau=\rho(\nu d\langle v_x \rangle/dy - \langle \delta v_x \delta
v_y\rangle)$ is constant in the shearwise direction (this follows
from the stationarity of the mean flow). Away from the wall $\tau
\simeq - \rho\langle \delta v_x \delta v_y\rangle$ whereas close
to the wall $\tau\simeq\rho \nu d\langle v_x \rangle/dy$;
consequently the Reynolds stress is usually normalized to
$\tau/\rho$, velocities to $v_{\tau}\equiv ({\tau/\rho})^{1/2}$,
and on has $v_M\simeq v_{\tau}$ in the bulk of the flow. The value
of $v_{\tau}$ for this simulation can be obtained in the following
way. Note first that Fig.~(1a) of their paper also displays
$Re_{\tau}^{-1}d\langle v_x \rangle/dy$ where, for their
simulation, $Re_{\tau}=v_{\tau}h/\nu= 82$, and where $v_x$ is
normalized to $v_{\tau}$ and $y$ to the walls half-distance $h$;
this quantity is equal to $1$ in the immediate vicinity of the
wall. On the other hand, the value of the velocity gradient near
the wall in units of $2U_w/h= \Delta v_x/h$ can be deduced from
Fig.~4 of \citet{Bech95} which relates to the same simulation. The
comparison of these two measures of the same quantity yields the
required value of $v_{\tau}/2U_w$.}, one finds $v_M\simeq \Delta
v_x/30$ for this simulation, which is nearly identical to the
estimate deduced from Eq.~(\ref{vm}). Even if one takes into
account the fact that the value of $\Delta v_x$ which is relevant
for the bulk of the flow is smaller than the one adopted in
Eq.~(\ref{vm}) by a factor $\sim 4$, the two estimates of $v_M$
still agree within a factor of $\sim 3$. Another estimate of the
same quantity for non-rotating free shear flows is obtained from
the representation\footnote{Note that $v_M$ characterizes
turbulent transport, and that $\langle \delta v_x \delta
v_y\rangle$ is usually smaller than the magnitude of velocity
fluctuations by a factor of a few.} of $\langle \delta v_x \delta
v_y\rangle$ in Fig.~14 (at $Q=0$) of \citet{Bid92}, and gives
$v_M\sim \Delta v_x/10$, which differs from the order of magnitude
estimate quoted above by a factor of $\sim 4$. Some indication of
the value of $l_M$ can also be extracted from Fig.~2 of
\citet{Bech95}, which shows the power spectra in the shearwise and
spanwise directions. However, the box size in these directions is
large in terms of $h$, and the resolution of the simulation does
not allow the authors to really reach the inertial part of the
turbulent spectrum. This is particularly noticeable for $k_x$
spectra in the middle of the flow (displayed in the $y=82$
quadrant of this figure), which are nearly flat down to $k_x
h\simeq 10$, and drop precipitously for larger values of $k$
because of numerical dissipation, as the limit resolution of the
simulation is reached. The $k_z$ spectra behave sensibly better,
more probably because the box is 2.5 times smaller in this
direction, and show some indications that an inertial spectrum
tries to develop for $8\lesssim k_z h\lesssim 30$. Because
$h=\Delta y/2$, this suggests that $l_M$ in this simulation is at
most within a factor of $\sim 3$ of the order of magnitude
estimate deduced from Eq.~(\ref{lm}). Note in passing that, for
Couette flows, the inertial spectrum does not need to be resolved
in order for turbulence to be observed in numerical simulations;
this is related to the existence of a large scale nonlinear
mechanism which sustains turbulence, as mentioned in section
\ref{plcou}, and which is most likely at the origin of the more or
less flat part of the spectra at large scales.

\clearpage

\clearpage

\clearpage

\end{document}